\documentclass[%
reprint,
preprintnumbers,
 amsmath,amssymb,
 aps,
]{revtex4-2}

\usepackage{graphicx}
\usepackage{dcolumn}
\usepackage{bm}
\usepackage[dvipsnames]{xcolor}

\usepackage[mathlines]{lineno}

\usepackage{physics}

\begin{document}

\preprint{TU-1265, KEK-QUP-2025-0015, MIT-CTP/5900}

\title{Detecting dark matter using optically trapped Rydberg atom tweezer arrays}

\author{
  So Chigusa$^{(a)}$,
  Taiyo Kasamaki$^{(b)\dagger}$,
  Toshi Kusano$^{(c)}$,
  Takeo Moroi$^{(b,d)}$,\\
  Kazunori Nakayama$^{(e,d)}$,
  Naoya Ozawa$^{(c)}$,
  Yoshiro Takahashi$^{(c)}$,
  Atsuhiro Umemoto$^{(d)}$,
  Amar Vutha$^{(f)}$
}

\affiliation{
$^{(a)}$Center for Theoretical Physics - a Leinweber Institute, Massachusetts Institute of Technology, Cambridge, MA 02139, USA
}

\affiliation{
$^{(b)}$Department of Physics, The University of Tokyo, Tokyo 113-0033, Japan
}

\affiliation{
$^{(c)}$Department of Physics, Graduate School of Science, Kyoto University, Kyoto 606-8502, Japan
}

\affiliation{
$^{(d)}$QUP (WPI), KEK, Tsukuba, Ibaraki 305-0801, Japan
}

\affiliation{
$^{(e)}$Department of Physics, Tohoku University, Sendai 980-8578, Japan
}

\affiliation{
$^{(f)}$Department of Physics, University of Toronto, Toronto ON M5S 1A7, Canada
}

\begin{abstract}

A new scheme for detecting wave-like dark matter (DM) using Rydberg atoms is proposed. Recent advances in trapping and manipulating Rydberg atoms make it possible to use Rydberg atoms trapped in optical tweezer arrays for DM detection. We propose to prepare a large ensemble of Rydberg atoms and to observe the excitations between Rydberg states by the DM-induced effective electric field. A scan over DM mass is enabled with the use of the Zeeman and diamagnetic shifts of energy levels under an applied external magnetic field. Taking dark-photon DM as an example, we demonstrate that our proposed experiment can have high enough sensitivity to probe previously unexplored regions of the parameter space of dark-photon coupling strengths and masses.

\end{abstract}

\maketitle

\noindent
$^{\dagger}${\it Corresponding author.}

\vspace{1mm}
\noindent
\underline{\it Introduction}: 
The origin of dark matter (DM) remains a long-standing mystery in modern cosmology and particle physics (see, e.g., \cite{ParticleDataGroup:2024cfk, Cirelli:2024ssz}). While the existence of DM has been confirmed through various astrophysical and cosmological observations, its particle-physics properties are still largely unknown. To gain a deeper understanding of the nature of DM, direct detection is essential. From this perspective, many experiments have been conducted or proposed; however, no conclusive evidence has been found so far.

One major challenge in the direct detection of DM is the uncertainty in its mass, which could range from $10^{-22}$\,eV to $10^{19}$\,GeV (or even higher). The strategy for the direct detection strongly depends on the mass of DM. If the mass is above approximately 1\,eV, DM behaves like a particle, and its detection may be possible via scattering with nuclei or electrons. In contrast, if DM is lighter than $\sim 1\, {\rm eV}$, DM exhibits wave-like behavior, and detection may involve the excitation of photons in a cavity or other excitations.

We focus on the direct detection of light wave-like DM. The conventional approach is through the use of resonant cavities~\cite{Sikivie:1983ip,Bradley:2003kg}, where DM is expected to excite photons within the cavity. Although these experiments have demonstrated high sensitivity to light DM candidates, such as axions and dark photons, no detection has been achieved so far. In light of this, it is important to explore new methods for detecting light DM, especially those that can cover a broader mass range.

Recently, proposals have emerged that aim to detect light DM using qubit-type sensors \cite{Dixit:2020ymh, Chen:2022quj, Chen:2023swh,Agrawal:2023umy,Ito:2023zhp,Chigusa:2023roq,Gue:2023jcd, Graham:2023sow, Engelhardt:2023qjf,Chen:2024aya,Braggio:2024xed,Chigusa:2024psk,Dong:2025mdk,Chen:2025tgj,Zhao:2025thg,Nakazono:2025tak,Bodas:2025vff,Banerjee:2025jss,Fukuda:2025afi}. These quantum sensors are sensitive to oscillating electric and/or magnetic fields, which can be induced by the oscillating DM fields. One potential advantage of these quantum sensors is tunability of their excitation energies, enabling DM searches over a wider mass range including those which are not covered by conventional cavity experiments \cite{ADMX:2025vom}. Given the rapid advancements in quantum sensing technologies, it is both timely and important to investigate their potential for the direct detection of light DM.

In this Letter, we propose a new method for detecting wave-like DM using Rydberg atoms. (For early attempts of using Rydberg atoms for DM detection, see \cite{Matsuki:1990mf, Ogawa:1996dr}.) Rydberg atoms are atoms with a highly excited valence electron, and their excitation energies are easily controlled by an external electric and/or magnetic field. Furthermore, because the transition rate between Rydberg states scales as $\sim O(\nu^4)$ with an effective principal quantum number $\nu$ (see, e.g., \cite{Gallagher:RydbergAtoms}), the sensitivity to DM-induced excitation is significantly enhanced. Our protocol involves preparing a large ensemble of Rydberg atoms in optical tweezer arrays (OTAs) and searching for DM-induced transitions between Rydberg states. By utilizing remarkable features of Rydberg atoms, the sensitivity to dark-photon DM could reach a parameter region unexplored before. In particular, we show that the search with Rydberg atoms can cover the dark-photon mass of $O(0.1)\, {\rm meV}$, which is challenging to access with conventional methods using cavities.

\vspace{1mm}
\noindent
\underline{\it Experimental setup}: We first introduce our proposal. Inspired by recent advances in manipulating Rydberg atoms~\cite{Bernien:2017, Chen:2023XY,Ma:2023high,Cao:2024multi,Finkelstein:2024,Bluvstein:2025ped}, we focus on an OTA platform~\cite{KaufmanNi:2021} as the experimental setup for DM detection.

\begin{figure}[t]
  \centering
  \includegraphics[width=0.95\linewidth]{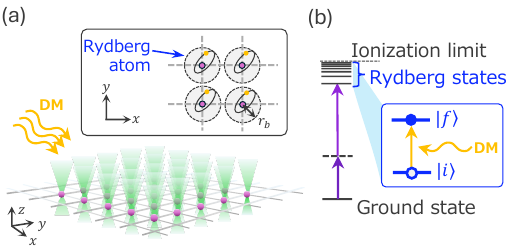}
  \caption{Schematic overview of the experiment. (a) Rydberg atoms trapped in OTAs. The tweezer array is sufficiently sparse that the site separation is larger than the Rydberg blockade radius $r_b$. Each green cone indicates the tightly focused laser to trap a Rydberg atom. (b) Energy levels of a Rydberg atom and the excitation due to the DM-induced effective electric field. The Rydberg state is prepared by exciting the atoms from the ground state to the highly excited state via two-photon excitation (purple arrows). The dashed line indicates the intermediate state for the two-photon excitation.}
  \label{fig:overview}
\end{figure}

In this system, individual neutral atoms are captured in an array of tightly focused laser beams and trapped for a minute-scale duration in an ultra-high vacuum environment. This duration is sufficiently long compared to the DM detection time (around 1~ms or shorter), ensuring that atom loss from the trap does not adversely affect the measurement. While the vacuum system surrounding OTAs is typically placed at room temperature, cryogenic vacuum systems have also become available~\cite{Schymik:2021,Pichard:2024,Zhang:2025}, which enables the suppression of blackbody radiation (BBR). In this study, we evaluate the DM sensitivity under two distinct conditions: a room-temperature (300 K) environment and a cryogenic environment cooled to 4 K (see later). By utilizing devices such as acousto-optic deflectors and spatial light modulators, the spatial configuration of the array can be freely designed~\cite{Nogrette:2014}. 
This facilitates placing Rydberg atoms at positions separated by more than the blockade radius $r_b$, the characteristic distance below which van der Waals or dipole-dipole interactions between Rydberg atoms become significant and multiple Rydberg excitations are suppressed.
While atoms in Rydberg states experience a repulsive force from the trap due to the ponderomotive force acting on the excited electron, alkaline-earth-like atoms in Rydberg manifolds are still trappable owing to the unexcited valence electron~\cite{Wilson:2022}. This enables an experiment utilizing a large number of well-isolated Rydberg atoms almost pinned to their initial positions (Fig.~\ref{fig:overview}(a)). The Rydberg atoms can be state-selectively ionized by a pulsed electric field, and the formed ions will subsequently be driven out of the trap and detected with an electric-charge-sensitive detector such as a microchannel plate. In a typical experimental setup, the atoms are initially loaded into a magneto-optical trap (MOT) and subsequently loaded into the OTA by overlapping the trapping lasers with the MOT atomic cloud. The atoms are prepared in their Rydberg states by extinguishing the MOT and applying an excitation laser. 

In the present analysis, we consider a class of wave-like DM that induces an oscillating effective electric field, which can drive transitions between Rydberg states. The observation of such transitions could be interpreted as a possible signal of wave-like DM.

\begin{figure}[t]
  \centering
  \includegraphics[width=0.95\linewidth]{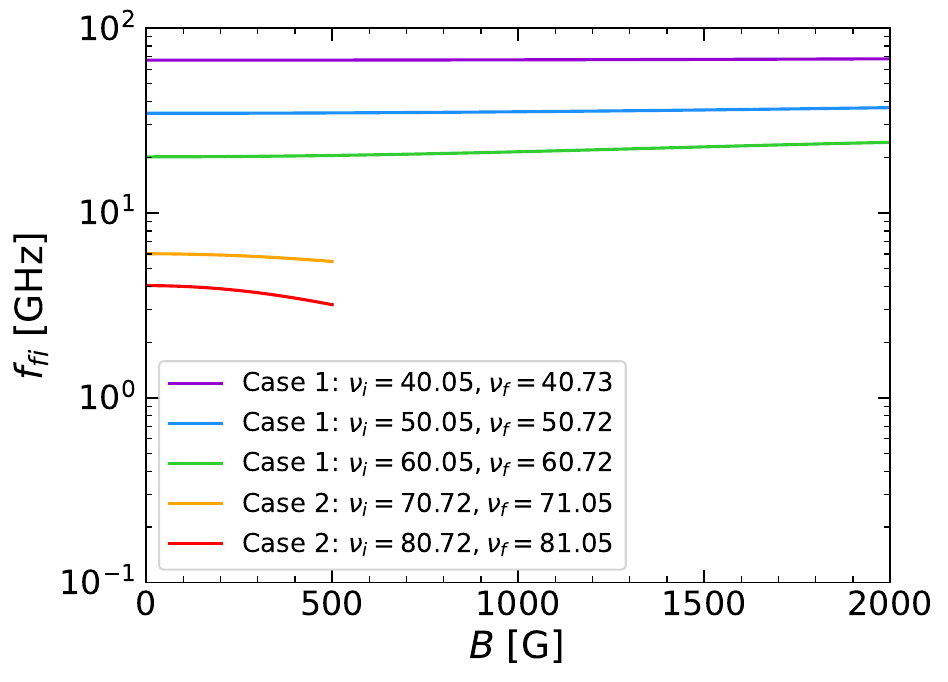}
  \caption{Resonance frequency $f_{fi}$ as a function of the external magnetic field $B$. The upper three lines are for Case 1 (with $\bar{\nu}=40$, $50$, and $60$), and the lower two lines are for Case 2 (with $\bar{\nu}=70$ and $80$), from above. For Case 2, data are shown only for $B \leq 500\,\mathrm{G}$ due to numerical instability at higher fields.}
  \label{fig:ffi}
\end{figure}

For the DM detection, we consider the following experimental setup. We use ytterbium (Yb) Rydberg atoms trapped in OTAs under a static magnetic field (the total number of Rydberg atoms is denoted by $n_{\rm Ryd}$). The external magnetic field is varied to scan over DM masses; we consider a magnetic field of $O(1)\ {\rm kG}$, which yields a frequency scan range of $O(1-10)\, {\rm GHz}$ (see Fig.\ \ref{fig:ffi}). Such a strong magnetic field does not hinder the readout quality of the Yb Rydberg states \cite{Muniz:2024}. Using the trapped Rydberg atoms, the following measurement cycle is repeated: (i) state preparation, in which all Rydberg atoms are initialized in a specific state $\ket{i}$; (ii) exposure to the DM-induced effective electric field for a finite duration; and (iii) readout of the Rydberg state. Then, the number of Rydberg atoms found in a target state $\ket{f}$ is counted and denoted as $N_{\rm excite}$. If $N_{\rm excite}$ is large enough compared to the noise level, we may claim the discovery of the wave-like DM. Using this counting-based approach, we demonstrate that the proposed method can achieve sensitivity to certain DM parameter regions yet to be explored. A schematic overview of the experiment is shown in Fig.\ \ref{fig:overview}.

To make our discussion simple and concrete, we consider the Rydberg states of $^{174}$Yb for DM detection. Its electron configuration at the ground state is [Xe]4f$^{14}$6s$^{2}$.  One of the electrons in the 6s orbit can be highly excited, realizing a Rydberg state, on which we focus hereafter. In the absence of external magnetic and electric fields, the energy eigenstates are labeled by four quantum numbers: $\nu$ (effective principal quantum number, with which the energy of the electron is given by $-R/\nu^2$ with $R\simeq 13.6\ {\rm eV}$), $\ell$ (orbital angular momentum), $F$ (total angular momentum), and $m$ (the $z$-component of the total angular momentum). The Rydberg state with fixed  $\nu$, $\ell$, $F$, and $m$ is denoted as $\ket{Q}\equiv\ket{\nu, \ell, F, m}$.  

\vspace{1mm}
\noindent
\underline{\it Excitation and deexcitation of Rydberg atoms}: To estimate the sensitivity, we consider both excitation and deexcitation processes of Rydberg atoms induced by the DM-induced effective electric field and by background BBR. In this Letter, we use the natural unit, i.e., $\hbar=c=1$.

We focus on the dynamics of a highly excited electron in a Rydberg atom. The effective Hamiltonian is described in the following form:
\begin{align}
  \hat{H} = \hat{H}_0 + \hat{H}_{\rm int}.
  \label{H_tot}
\end{align}
Here, $\hat{H}_0$ denotes the ``free'' part, which includes the potential due to the nuclear core and inner-core electrons, as well as the effects of external magnetic or electric fields (if present). $\hat{H}_{\rm int}$ describes the interaction between the electron and the DM-induced effective electric field and is given by
\begin{align}
  \hat{H}_{\rm int} = e \vec{E} (t) \cdot \hat{\vec{r}},  
\end{align}
where $\hat{\vec{r}}\equiv (\hat{x},\hat{y},\hat{z})$ denotes the position operator of the electron. The DM-induced effective electric field oscillates with angular frequency equal to the DM mass $m_X$, and $\vec{E}$ is given by
\begin{align}
  \vec{E} \equiv \bar{E} \vec{n} \sin (m_X t + \phi),
  \label{vecE}
\end{align}
where $\bar{E}$ is the amplitude while $\vec{n}$ is the unit vector parallel to $\vec{E}$. (If a conducting wall surrounds the Rydberg atom, the ordinary electric field generated at the surface may cancel the DM-induced effective electric field. However, if the distance between the wall and the atom is much larger than $m_X^{-1}$, such cancellation is avoided. In our analysis, we neglect the effect of the conducting wall, assuming that the system volume is sufficiently large to justify this approximation and to derive a representative sensitivity; see, e.g., Ref.~\cite{Chen:2022quj}.) We classify the states by the eigenstates of $\hat{H}_0$, treating $\hat H_{\rm int}$ perturbatively.

Let us consider the transition from the Rydberg state $\ket{i}$ to $\ket{f}$ (with energies $E_i$ and $E_f$, respectively). The transition process is governed by the matrix element $\vec{r}_{fi} \equiv \mel{f}{\hat{\vec{r}}}{i}$, with which the transition rate (i.e., the excitation or deexcitation probability per unit time) due to the DM-induced effective electric field, denoted as $\gamma_{i\rightarrow f}^{\mathrm{(DM)}}$, can be evaluated. In addition, even without DM, the transition may occur. Particularly, with BBR around the Rydberg atom, we expect that the stimulated and spontaneous emission (for $E_i>E_f$) or the transition with absorbing the BBR (for $E_i<E_f$) is the dominant noise source; the transition rate of such a process is denoted as $\gamma_{i\rightarrow f}^{\mathrm{(rad)}}$. Calculations of $\gamma_{i\rightarrow f}^{\mathrm{(DM)}}$ and $\gamma_{i\rightarrow f}^{\mathrm{(rad)}}$ are detailed in End Matter.

\vspace{1mm}
\noindent
\underline{\it Dark-photon DM}: To examine the sensitivity of the DM detection experiment of our proposal, we consider a concrete example, i.e., dark photon, which is one of well-motivated DM candidates~\cite{Graham:2015rva,Ema:2019yrd,Long:2019lwl,Kitajima:2022lre,Kitajima:2023fun}. The dark photon is a massive vector field having kinetic mixing with the ordinary photon. In the mass-eigenstate basis of the photon $A_\mu$ and dark photon $X_\mu$, where the kinetic-mixing term is eliminated, the Lagrangian for the electron field $\psi$ is given by
\begin{align}
  \mathcal{L} \ni 
  \bar{\psi} \left[
    (i\partial_\mu + e A_\mu + \epsilon e  X_\mu)\gamma^\mu - m_e
    \right]   
    \psi,
    \label{Lagrangian}
\end{align}
where $e$ is the electric charge, $m_e$ is the electron mass, and $\epsilon$ is the dimensionless coupling strength of the dark photon. The dark-photon vector potential around the Earth can be approximated as
\begin{align}
  \vec{X} = \bar{X} \vec{n} \cos  (m_X t + \phi),
  \label{Xosc}
\end{align}
where $\bar{X}$ is the amplitude of the oscillation and $m_X$ is the dark-photon mass.  The amplitude is related to the local DM density as $\rho_{\rm DM} = \frac{1}{2} m_X^2 \bar{X}^2$. As shown in Eq.\ \eqref{Lagrangian}, dark photons couple to charged particles in a similar way to ordinary photons. The dark-photon-induced effective electric field is evaluated as $\vec{E}= -\epsilon \dot{\vec{X}}$, yielding Eq.\ \eqref{vecE} with $\bar{E} = 
  \epsilon m_X \bar{X}
  = \epsilon \sqrt{2 \rho_{\rm DM}}$.
Then, we obtain
\begin{align}
  \gamma_{i\rightarrow f}^{\rm (DM)} =
  \frac{2\pi}{3} \alpha \epsilon^2 \rho_{\rm DM} |\vec{r}_{fi}|^2 \tau 
  W((m_X-\omega_{fi}) \tau),
  \label{gammaDM}
\end{align}
where $W(\mu)\equiv 2\mu^{-2}(1 - \cos \mu)$ and $\omega_{fi} \equiv 2\pi f_{fi} \equiv |E_f-E_i|$ is the transition energy. (See End Matter for details.) We emphasize that the transition rate is enhanced in the resonance limit, i.e., $\omega_{fi}\rightarrow m_X$. We set the exposure time to the coherence time of the system $\tau$; we take $\tau = \mbox{min}(\tau_{\rm Ryd}, \tau_{\rm DM})$, where $\tau_{\rm Ryd}$ is the shorter of the lifetimes of the Rydberg states $\ket{i}$ and $\ket{f}$, while $\tau_{\rm DM}\sim \frac{2\pi}{m_X v_{\rm DM}^2}$ (with $v_{\rm DM}\sim 10^{-3}$ being the typical velocity of the DM) is the coherence time of the DM. Considering Rydberg states with their (effective) principal quantum numbers $\sim \nu$, $|\vec{r}_{fi}|\sim O(a_{\rm B}\nu^2)$ (with $a_{\rm B}$ being the Bohr radius) \cite{Gallagher:RydbergAtoms}. Thus, we expect the transition rates to be further enhanced when more highly excited Rydberg atoms are used.

\vspace{1mm}
\noindent
\underline{\it Dark-photon DM detection with Rydberg atoms}: We now examine the sensitivity of dark-photon DM searches using Rydberg atoms. 
%
%
%
As mentioned earlier, the transition rate of a Rydberg atom is enhanced in the resonance limit. Therefore, two Rydberg states whose energy separation matches the DM mass should be considered. Since the DM mass is unknown, we need to scan over possible DM masses by varying the energy separation. To achieve this, we propose utilizing the Zeeman and diamagnetic effects~\cite{Neukammer:1984} to tune the energy levels of Rydberg states by applying an external magnetic field. In the presence of the external magnetic field, the energy eigenstates are expressed as linear combinations of states with different quantum numbers:
\begin{align}
  \ket{\Psi}_B =
  \sum_Q
  U_{\Psi, Q}
  \ket{Q}.
  \label{state*}
\end{align}
Here and hereafter, the states with the subscript ``$B$'' denote energy eigenstates with the external magnetic field. In particular, the state $\ket{\nu, \ell, F, m}_B$ denotes the energy eigenstate that has the largest overlap with the basis state $\ket{\nu, \ell, F, m}$; specifically, for $\ket{\Psi}_B = \ket{\nu, \ell, F, m}_B$, the overlap $|U_{\Psi, \{\nu, \ell, F, m\}}|$ is the largest among all $|U_{\Psi', \{\nu, \ell, F, m\}}|$ as $\Psi'$ varies. We use the {\tt rydcalc} package~\cite{rydcalc} to compute quantities necessary to evaluate the transition rate. (For details, see End Matter.)

By selecting various combinations of $\ket{i}$ and $\ket{f}$, the transition energy $\omega_{fi}$ can be varied, enabling the exploration of different DM mass ranges. The matrix element $\vec{r}_{fi}$ tends to become more suppressed as $|\nu_i - \nu_f|$ grows (where $\nu_i$ and $\nu_f$ are effective principal quantum numbers of $\ket{i}$ and $\ket{f}$, respectively). To avoid this suppression while scanning a broad range of DM masses, we focus on transitions between states satisfying $|\nu_i - \nu_f| \lesssim 1$. 
A comprehensive and systematic investigation of other possibilities is left for future work.

We consider the following Rydberg states of $^{174}$Yb as representative examples of the initial and final states:
\begin{align*}
  \mbox{Case 1:}\,&
  \ket{i} = \ket{\bar{\nu}+\delta\nu_1, 1, 1, 0}_B,\,
  \ket{f} = \ket{\bar{\nu}+\delta\nu_0, 0, 0, 0}_B,
  \\
  \mbox{Case 2:}\,&
  \ket{i} = \ket{\bar{\nu}\!+\!\delta\nu_0, \!0, \!0, \!0}_B,\,
  \ket{f} = \ket{\bar{\nu}\!+\!1\!+\!\delta\nu_1, \!1, \!1, \!0}_B,
\end{align*}
where $\bar{\nu}$ is an integer in the range $30 \leq \bar{\nu} \leq 60$ for Case 1 and $70 \leq \bar{\nu} \leq 88$ for Case 2 and $\delta\nu_0$ and $\delta\nu_1$ are the fractional parts of the effective principal quantum numbers ($\delta\nu_0 \simeq 0.72$ and $\delta\nu_1 \simeq 0.05$ \cite{Peper:2024gmw}). In both cases, $E_f > E_i$ (for the magnetic field $B$ adopted in the present analysis). With such setups, we can access dark-photon DM with $m_X\lesssim O(100)\, \mu{\rm eV}$, for which the de Broglie wavelength $1/m_X v_{\rm DM}$ is $O(1)\,{\rm m}$ or longer.

In Fig.\ \ref{fig:ffi}, we show $f_{fi}$ as a function of the external magnetic field $B$ for Case 1 with $\bar{\nu}=40$, $50$, and $60$ as well as for Case 2 with $\bar{\nu}=70$ and $80$. For Case 1, we can see that $f_{fi}$ is of $O(10)\,\mathrm{GHz}$ and varies by several percent or more when $B \sim O(1000)\,\mathrm{G}$. By appropriately tuning $B$ as well as $\bar{\nu}$, it is possible to continuously scan the DM mass in the $O(100)\,\mu\mathrm{eV}$ range, which conventional cavity-based experiments hardly access. It is also notable that much smaller values of $f_{fi}$, significantly below $10\,\mathrm{GHz}$, are achievable, as exemplified by Case 2. 

Both the DM-induced signal and the background noise contribute to $N_{\mathrm{excite}}$. The expected number of signal events is given by
\begin{align}
  N_{\mathrm{signal}} = n_{\mathrm{Ryd}}\, \gamma_{i\rightarrow f}^{\mathrm{(DM)}}\, t_{\mathrm{bin}},
\end{align}
where $t_{\mathrm{bin}}$ denotes the total integration time per frequency bin. Notice that, because the resonance condition should be met, we need to specify the final state. We expect that the duration of a single measurement cycle can be comparable to the exposure time (which is taken to be the coherence time $\tau$), owing to the recent efforts to minimize the sequence time and developments that enable continuous loading\ \cite{Norcia2024-lp,Chiu2025-so,Li2025-yc}. Also, the atoms remaining in $\ket{i}$ after the exposure to the DM-induced field may be brought back to their ground states via optical means, allowing the sequence to be repeated without state preparation. Therefore, each bin consists of $\sim\frac{t_{\mathrm{bin}}}{\tau}$ repeated measurement cycles. Errors in state preparation could be minimized by applying an ionization electric field pulse immediately after the preparation to drive away the atoms initialized in the undesired state (see Ref.\ \cite{Tada:2002}) and carefully engineering the pulse shape of the optical excitation to the Rydberg state\ \cite{Shi2021-xt}. Thus, the BBR-induced absorption is expected to be the dominant source of the background noise. In this case, presuming that the readout procedure involves ionizing all states above a threshold energy $E_{\rm th}$, the associated number of noise events is evaluated as
\begin{align}
  N_{\mathrm{noise}} = n_{\mathrm{Ryd}}
  \sum_{\alpha}
  \gamma_{i\rightarrow\alpha}^{\mathrm{(rad)}}\, 
  \theta (E_\alpha - E_{\rm th})\, t_{\mathrm{bin}}.
\end{align}
We set $E_{\rm th}=E_f$ (so that $E_i<E_f$ is required to identify the excitation signal).


If $N_{\mathrm{excite}}$ is observed to be sufficiently large compared to the noise level in a specific frequency bin, it may indicate the Rydberg atom excitation due to DM.
We evaluate the sensitivity to dark-photon DM, adopting the following criterion to define sensitivity: $N_{\mathrm{signal}} >2\sqrt{N_{\mathrm{noise}}}$.
To determine the appropriate scan step, we approximate $\gamma_{i\rightarrow f}^{\mathrm{(DM)}}$ as follows: $\gamma_{i\rightarrow f}^{\mathrm{(DM)}}=\gamma_{i\rightarrow f}^{\mathrm{(DM)}}(\omega_{fi}=m_X)$ if $\omega_{fi} - \tau^{-1} \leq m_X \leq \omega_{fi} + \tau^{-1}$ and $\gamma_{i\rightarrow f}^{\mathrm{(DM)}}=0$ otherwise. The scan step of the magnetic field is then chosen such that the width of each frequency bin becomes $2\tau^{-1}$. 

\begin{figure}[t]
  \includegraphics[width=0.95\linewidth]{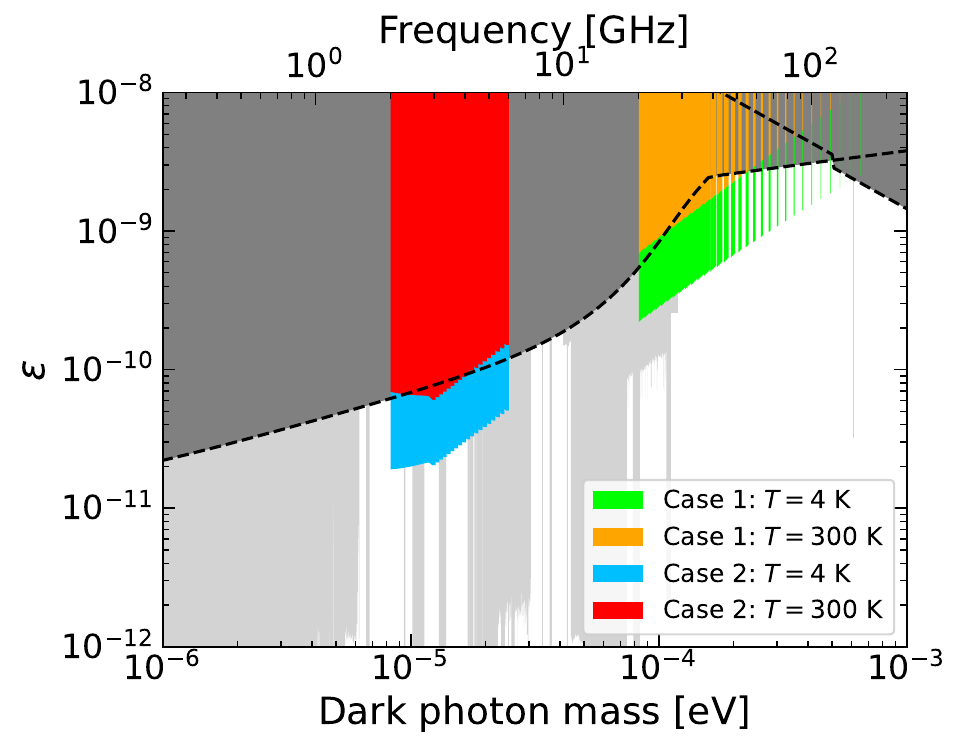}
  \caption{Expected sensitivity to dark-photon DM, assuming $t_{\rm bin} = 10\,\mathrm{s}$, $n_{\mathrm{Ryd}} = 10^3$, and $\rho_{\rm DM} = 0.45\,\mathrm{GeV/cm^3}$. The external magnetic field is varied within $0 \leq B \leq 2000\,\mathrm{G}$ for Case 1 and $0 \leq B \leq 500\,\mathrm{G}$ for Case 2. For Case 1 (Case 2), sensitivity regions for $\bar{\nu} = 30-60$ ($\bar{\nu} = 70-88$) are shown from right to left; the regions overlap for large enough $\bar{\nu}$. The dark gray region above dashed lines is excluded by cosmological or astrophysical considerations \cite{McDermott:2019lch,Vinyoles:2015aba,An:2020bxd}, while the light gray regions are excluded by haloscope experiments \cite{Knirck:2018ojz, Tomita:2020usq, Brun:2019kak, Levine:2024noa, Nguyen:2019xuh, Cervantes:2022yzp, DOSUE-RR:2023yml, An:2022hhb, Cervantes:2022gtv, Ramanathan:2022egk, Bajjali:2023uis, SHANHE:2023kxz, BREAD:2023xhc, MADMAX:2024jnp, ADMX:2025vom, HAYSTAC:2024jch, CAPP:2024dtx, Adair:2022rtw, TASEH:2022vvu, Kang:2024slu, McAllister:2022ibe, QUAX:2024fut, Schneemann:2023bqc, APEX:2024jxw} or quantum cyclotron \cite{Fan:2022uwu}. (We use the dataset provided by \cite{Caputo:2021eaa}.)}
  \label{fig:sensitivity}
\end{figure}

In Fig.\ \ref{fig:sensitivity}, we present the expected sensitivity for Case 1 with $30\leq\bar{\nu}\leq 60$ as well as for Case 2 with $70\leq\bar{\nu}\leq 88$. Two representative temperatures, $T = 4$ and $300\,\mathrm{K}$, are considered and the magnetic field is varied within $0 \leq B \leq 2000\,\mathrm{G}$ for Case 1 and $0 \leq B \leq 500\,\mathrm{G}$ for Case 2. Here, we take $n_{\mathrm{Ryd}} = 10^3$ and $t_{\mathrm{bin}} = 10\,\mathrm{s}$; the sensitivity to $\epsilon$ scales as $(n_{\rm Ryd} t_{\rm bin})^{-1/4}$. In the calculation for Fig.\ \ref{fig:sensitivity}, $\tau$ is set by $\tau_{\rm DM}$ except in Case 2 at $T=300\ {\rm K}$ for $m_X\lesssim 12\ \mu{\rm eV}$. In Case 1, we can probe dark-photon DM with $m_X\sim O(100)\,\mu\textrm{eV}$. In particular, using the Rydberg states with $50\lesssim\bar{\nu}\lesssim 55$ and properly designing the scan ladder with changing $\bar{\nu}$ and $B$, we can continuously scan the mass range of $100\lesssim m_X\lesssim 150\, \mu\mathrm{eV}$ (with $0 \leq B \leq 2000\,\mathrm{G}$); the total amount of time for scanning such a DM mass range is $\sim 1\,\mathrm{year}$. If we can utilize $10^4$ atoms, which is reasonable considering the recent advances in OTA technologies \cite{Schlosser:2019hhm,Pause:2023pao,Manetsch:2024lwl}, the same sensitivity is obtained in a month or so (with shortening $t_{\rm bin}$). It is also notable that a wider mass range can be continuously probed if a stronger magnetic field is available. In Case 2, we may probe a much smaller DM mass range. The DM search of our proposal may reach an unexplored region of parameter space even with $O(10^2)$ atoms even at room temperature. Conducting the experiment at lower temperatures~\cite{Schymik:2021,Pichard:2024,Zhang:2025}  improves the sensitivity.

\vspace{1mm}
\noindent
\underline{\it Summary}: We have proposed a new method of detecting wave-like DM using Rydberg atoms. Using the fact that a class of wave-like DM generates an oscillating effective electric field, we may perform the DM detection experiment by looking for excitation processes of the Rydberg atoms due to the DM-induced effective electric field. In particular, relying on recent progresses in OTA technologies on neutral atoms, we consider the possibility of using trapped Rydberg atoms for DM detection. Considering the dark-photon DM as an example, we have shown that the experiment with Rydberg atoms can probe parameter regions that are currently unexplored. Scan over the DM mass is possible with the help of the Zeeman and diamagnetic effects. 

Our procedure offers unique features and advantages over previous proposals for DM detection using Rydberg atoms, which include the use of Rydberg atoms as microwave photon detectors coupled to axion cavities \cite{Matsuki:1990mf, Ogawa:1996dr, Gue:2023jcd, Graham:2023sow}, as well as the detection of axion DM using Rydberg atomic gases \cite{Engelhardt:2023qjf}. (For gravitational wave detection using Rydberg atoms, see also~\cite{Kanno:2023whr}.) We employ trapped Rydberg atoms.
Furthermore, compared to cavity haloscope experiments, scanning over the DM mass is more straightforward in our setup thanks to the Zeeman and diamagnetic shift; it does not require physical adjustment of cavity boundary conditions. In addition, the extensive techniques for manipulating Rydberg atoms, largely developed for quantum information processing, directly benefit our experiment. We may also utilize quantum features of Rydberg atoms to enhance detection sensitivity; quantum enhancement of the signal rate using entangled states is one such possibility \cite{Chen:2023swh, Chen:2025tgj, Bodas:2025vff, Fukuda:2025afi}.

\vspace{1mm}
\noindent
\underline{\it Acknowledgements}: This work is supported by the Simons Foundation (SC), the U.S. Department of Energy, Office of Science, the BNL C2QA award under grant Contract Number DESC0012704 (SUBK\#390034) (SC), JSPS KAKENHI Grants Nos.\ 23K22486 (TM), 24K07010 (KN), 24KJ0120 (NO) and 25K17412 (NO), and JST SPRING Grant No. JPMJSP2110 (TK), JST CREST Grant No.\ JPMJCR23I3(YT), MEXT Quantum Leap Flagship Program (MEXT Q-LEAP) Grant No.\ JPMXS0118069021 (YT), and JST Moon-shot R\&D Grant Nos.\ JPMJMS2268 (YT) and JPMJMS2269 (YT). AV and YT acknowledge JSPS International Research Fellowship No. S24151.

\bibliography{ref_arXiv_v2}

\pagebreak
\setcounter{equation}{0}
\setcounter{figure}{0}
\setcounter{table}{0}
\makeatletter
\renewcommand{\theequation}{E.\arabic{equation}}
\renewcommand{\thefigure}{E.\arabic{figure}}
\onecolumngrid
\begin{center}
\textbf{\large End Matter}
\end{center}
\twocolumngrid

In this End Matter, we explain our procedure to calculate the transition rates of the Rydberg atom. The calculation is based on the Hamiltonian given in Eq.\ \eqref{H_tot}. The interaction term is in the following form:
\begin{align}
  \hat{H}_{\rm int} = e \vec{E} (t)\cdot \hat{\vec{r}},
  \label{E:Hint}
\end{align}
with
\begin{align}
  \vec{E} (t) = \bar{E} \vec{n} \sin (m_X t + \phi).
  \label{E:vecE}
\end{align}

In the following, we analyze the evolution of the Rydberg state by decomposing it into a linear combination of the eigenstates of $\hat{H}_0$. With the Hamiltonian eigenstates $\ket{\alpha}$ (which are time-independent), satisfying
\begin{align}
  \hat{H}_0 \ket{\alpha} = E_\alpha \ket{\alpha},
\end{align}
we decompose the state as
\begin{align}
  \ket{\psi (t)} = \sum_{\alpha} C_\alpha (t) e^{-iE_\alpha t} \ket{\alpha}.
\end{align}

Evolution of the state can be analyzed by solving the Schr\"odinger equation:
\begin{align}
  i \frac{d}{dt} \ket{\psi (t)} = 
  (\hat{H}_0 + \hat{H}_{\rm int}) \ket{\psi (t)},
\end{align}
which gives
\begin{align}
  i \dot{C}_\alpha = 
  \sum_{\beta}
  e \bar{E} (\vec{n} \cdot \vec{r}_{\alpha\beta})
  e^{-i(E_\beta-E_\alpha)t}
  \sin( m_X t+\phi)
  C_\beta,
\end{align}
with the ``dot'' denoting the derivative with respect to time and $\vec{r}_{\alpha\beta}\equiv\mel{\alpha}{\hat{\vec{r}}}{\beta}$.  Because we are interested in the case that the state is initially $\ket{i}$ (i.e., $C_i(0)=1$) and that the effect of the interaction Hamiltonian can be perturbatively treated, at the leading order in the perturbation,
\begin{align}
  i \dot{C}_{f} \simeq
  2 \eta
  e^{-i(E_i-E_f)t} \sin(m_X t+\phi), 
\end{align}
where $\eta\equiv\frac{1}{2} e \bar{E} (\vec{n}\cdot \vec{r}_{fi})$.
Then we find
\begin{align}
  C_{f\neq i} \simeq &\,
  i\eta 
  e^{-i\phi} \frac{e^{-i(m_X + E_i-E_f)t}-1}{m_X + E_i-E_f}
  \nonumber \\ &\,
  + i\eta e^{i\phi} \frac{e^{i(m_X-E_i+E_f)t}-1}{m_X - E_i+E_f}.
  \label{E:C_fi}
\end{align}
In the resonance limit, i.e., $|m_X-\omega_{fi}|\ll m_X$ (with $\omega_{fi}\equiv|E_f-E_i|$), one of the terms in the right-hand side of Eq.\ \eqref{E:C_fi} dominates over the other. Neglecting the sub-dominant term, the transition probability from $\ket{i}$ to $\ket{f}$ at time $t$, which is given as $P_{fi} (t)=|C_{f}|^2$, is estimated as
\begin{align}
  P_{fi} (t) \simeq (\eta t)^2\, W((m_X-\omega_{fi})t),
  \label{E:Prob}
\end{align}
where
\begin{align}
  W (\mu) = \frac{2 (1 - \cos \mu)}{\mu^2}.
  \label{E:fnW}
\end{align}
The function $W(\mu)$ is peaked at $\mu=0$ (with $W(0)=1$) and decreases as $O(\mu^{-2})$ when $\mu\gg 1$; its half-maximum full width is $\sim 5.6$. Thus, for fixed $t$, the sensitivity to DM is maximized in the resonance limit, i.e., $\omega_{fi}\rightarrow m_X$. 

Importantly, the phase parameter $\phi$ in Eq.\ \eqref{E:vecE} depends on time. Such a time dependence can be approximated by a random reset of $\phi$ every DM coherence time $\tau_{\rm DM}$. We assume that the direction of the DM-induced effective electric field is randomized with the same time scale. In addition, the Rydberg state has a finite lifetime, which determines its coherence time. Because Eq.\ \eqref{E:Prob} is obtained by assuming the coherent evolution of the state with fixed $\phi$, it is applicable only for $t$ shorter than the coherence time of the system $\tau$. Then, the effective transition rate of the Rydberg atom, which is the transition probability per unit time averaged over measurement cycles (with each exposure time equal to $\tau$), is estimated as
\begin{align}
  \gamma_{i\rightarrow f}^{\rm (DM)} \equiv &\, 
  \frac{\langle P_{fi} (\tau)\rangle}{\tau}
  =
  \frac{\pi}{3} \alpha \bar{E}^2 |\vec{r}_{fi}|^2 \tau 
  W((m_X-\omega_{fi}) \tau),
  \label{E:gamma(DM)}
\end{align}
where $\alpha$ is the fine-structure constant and $\langle\cdots\rangle$ denotes the average over a number of measurement cycles (including the average over the direction of $\vec{n}$; for details, see \cite{Caputo:2021eaa}).  In our numerical calculation, (i) $\tau_{\rm Ryd}$ is set to the shorter of the lifetimes of the initial and final Rydberg states, computed using the {\tt Pairinteraction} package \cite{Weber2017} without considering external magnetic fields; and (ii) we take $v_{\rm DM}=10^{-3}$. We found that the lifetime of the Rydberg state for the case of our interest is $O(0.1-1)\ {\rm ms}$.

We can also estimate the transition rate for the absorption process of the BBR or the stimulated and spontaneous emission process. For this purpose, we replace the (effective) classical electric field in Eq.\ \eqref{E:Hint} with the corresponding operator; adopting the box normalization (with volume $V$), 
\begin{align}
  \hat{H}_{\rm int} = e \hat{\vec{E}} \cdot \hat{\vec{r}}.
\end{align}
Here, 
\begin{align}
  \hat{\vec{E}} = \frac{i}{\sqrt{2V}} \sum_{\vec{k},\lambda} k^{1/2}
  \left( 
  \hat{c}_{\vec{k},\lambda} \vec{\varepsilon}_{\vec{k},\lambda} e^{i\vec{k}\cdot\vec{x}}
  - \hat{c}_{\vec{k},\lambda}^\dagger \vec{\varepsilon}_{\vec{k},\lambda} 
  e^{-i\vec{k}\cdot\vec{x}}
  \right),
\end{align}
where $k\equiv |\vec{k}|$, $\lambda$ is the polarization index, and $\vec{\varepsilon}_{\vec{k},\lambda}$ is the polarization vector satisfying $\vec{k}\cdot\vec{\varepsilon}_{\vec{k},\lambda}=0$. In addition, $\hat{c}_{\vec{k},\lambda}$ and $\hat{c}_{\vec{k},\lambda}^\dagger$ are annihilation and creation operators satisfying $[\hat{c}_{\vec{k},\lambda}, \hat{c}_{\vec{k}',\lambda'}^\dagger]=\delta_{\vec{k},\vec{k}'}\delta_{\lambda,\lambda'}$.

\begin{figure}[t]
  \includegraphics[width=0.95\linewidth]{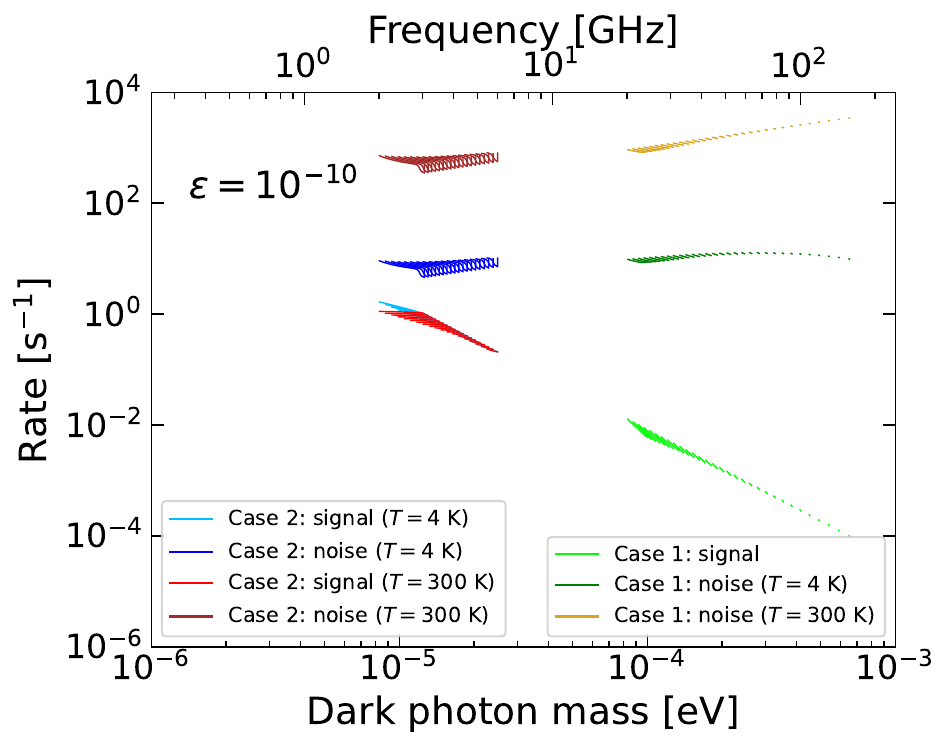}
  \caption{Signal (with $\epsilon=10^{-10}$) and noise rates as functions of the DM mass for Case 1 and Case 2, for $T=4$ and $300\, \mathrm{K}$. The external magnetic field is varied within $0\leq B \leq 2000\, \mathrm{G}$ for Case 1 and $0\leq B \leq 500\,\mathrm{G}$ for Case 2. Each line segment shows the result for a fixed value of $\bar{\nu}$; for Case 1, we take $\bar{\nu}=30$ (rightmost), $31$, $\ldots$, $59$, $60$ (leftmost); for Case 2, $\bar{\nu}=70$ (rightmost), $71$, $\ldots$, $87$, $88$ (leftmost). Note that, for Case 1, the signal rate does not depend on $T$, because $\tau$ is always determined by the DM coherence time.}
  \label{fig:rate}
\end{figure}

For the absorption process, we prepare the initial and final states  (denoted as $\ket{I}$ and $\ket{F}$, respectively) as the direct product of the state describing the Rydberg atom ($\ket{i}$ or $\ket{f}$) and the photon state:
\begin{align}
  \ket{I} = \ket{i} \otimes \ket{n_\gamma},~~~
  \ket{F} = \ket{f} \otimes \ket{n_\gamma-1}.
\end{align}
Here, the photon state is taken to be a number eigenstate (with fixed momentum and polarization), i.e., $\ket{n_\gamma} = \frac{1}{\sqrt{n_\gamma!}} (c_{\vec{k},\lambda}^\dagger)^{n_\gamma} \ket{0}$, with $n_\gamma$ being the number of photons (with the momentum $\vec{k}$ and the polarization $\lambda$). Decomposing the state as
\begin{align}
  \ket{\Psi (t)} = C_I (t) e^{-iE_I t} \ket{I} + C_F (t) e^{-iE_F t} \ket{F} + \cdots,
\end{align}
with $E_I=E_i+n_\gamma k$ and $E_F=E_f+(n_\gamma-1) k$, and using the initial condition $C_I(0)=1$ while $C_F(0)=0$, we find
\begin{align}
  \dot{C}_F (t) \simeq &\,
  \frac{k^{1/2}}{\sqrt{2V}} e
  (\vec{\varepsilon}_{\vec{k},\lambda}\cdot \vec{r}_{fi}) \sqrt{n_\gamma}
  e^{-i(E_i - E_f + k) t}.
\end{align}
It gives the transition rate of the absorption process of the photon with the momentum $\vec{k}$ and the polarization $\lambda$ as
\begin{align}
  \gamma (\vec{k},\lambda) =
  \frac{|C_F|^2}{t}
  =
  \frac{\pi}{V} e^2 k   (\vec{\varepsilon}_{\vec{k},\lambda}\cdot\vec{r}_{fi})^2 n_\gamma \delta (k+E_i-E_f),
  \label{E:gamma_k}
\end{align}
where the following relation is used:
\begin{align}
  \frac{1}{t} \left| \int_0^t dt' e^{i \Omega t'} \right|^2
  \xrightarrow{t\rightarrow\infty} 
  2 \pi \delta (\Omega).
\end{align}
For the case of stimulated and spontaneous emission, a similar calculation can be performed; the result is given by Eq.\ \eqref{E:gamma_k} with replacing $n_\gamma\rightarrow n_\gamma+1$ and $\delta (k+E_i-E_f)\rightarrow\delta (k-E_i+E_f)$.

Total transition rate is obtained by integrating over the momentum of the photon; replacing $n_\gamma$ by the distribution function, we obtain
\begin{align}
  \gamma_{i\rightarrow f}^{\rm (rad)} 
  = &\,
  \frac{4}{3} \alpha \omega_{fi}^3
  |\vec{r}_{fi}|^2
  \times
  \left\{
  \begin{array}{ll}
    f (\omega_{fi}) & :~ E_i<E_f
    \\[2mm]
    1+f (\omega_{fi}) & :~ E_i>E_f
  \end{array}
  \right. ,
\end{align}
where $f(|\vec k|)=(e^{|\vec{k}|/T}-1)^{-1}$ (with $T$ being the temperature).
%

In our analysis, the  {\tt rydcalc} package~\cite{rydcalc} is used for the calculation of the matrix elements. Note that {\tt rydcalc} provides the matrix elements $\mel{n, \ell, F, m}{\hat{\vec{r}}}{n', \ell', F', m'}$, whereas we are interested in transitions between the energy eigenstates under the influence of the external magnetic field (see Eq.~\eqref{state*}). The relevant matrix elements are expressed in terms of those provided by {\tt rydcalc} as
\begin{align}
  \vec{r}_{fi} =  
  \sum_{Q_1, Q_2} U_{f, Q_1}^* \, U_{i, Q_2} \mel{Q_1}{\hat{\vec{r}}}{Q_2}.
\end{align}

In Fig.\ \ref{fig:rate}, we show $\gamma_{i\rightarrow f}^{\mathrm{(DM)}}$ and $\gamma_{i\rightarrow f}^{\mathrm{(rad)}}$ as functions of the DM mass $m_X$, considering Case 1 with $30\leq\bar{\nu}\leq 60$ as well as Case 2 with $70\leq\bar{\nu}\leq 88$. Here, we adopt $T=4$ and $300\, \mathrm{K}$, set $\epsilon=10^{-10}$, and vary the magnetic field in the range $0 \leq B \leq 2000\,\mathrm{G}$ for Case 1 and $0 \leq B \leq 500\,\mathrm{G}$ for Case 2.

\end{document}